\documentclass[journal]{IEEEtran}
\usepackage{amsmath,amsfonts,graphicx,tabularx,multirow}
\usepackage{color}
\usepackage{algorithm} 
\usepackage{algorithmic}  
\usepackage[algo2e]{algorithm2e} 
\usepackage{bm,mathtools,upgreek}

\def\thline{\noalign{\hrule height 1.0pt}}

\DeclareMathOperator*{\argmin}{argmin}

\renewcommand{\vec}[1]{\bm{\mathrm{#1}}}

\title{Speaker-independent Speech Separation with Deep Attractor Network}

\author{Yi~Luo\thanks{Yi Luo and Zhuo Chen contributed equally to this work.}, Zhuo~Chen, and~Nima~Mesgarani}

\begin{document}

\maketitle


\begin{abstract}
Despite the recent success of deep learning for many speech processing tasks, single-microphone, speaker-independent speech separation remains challenging for two main reasons. The first reason is the arbitrary order of the target and masker speakers in the mixture (permutation problem), and the second is the unknown number of speakers in the mixture (output dimension problem). We propose a novel deep learning framework for speech separation that addresses both of these issues. We use a neural network to project the time-frequency representation of the mixture signal into a high-dimensional embedding space. A reference point (attractor) is created in the embedding space to represent each speaker which is defined as the centroid of the speaker in the embedding space. The time-frequency embeddings of each speaker are then forced to cluster around the corresponding attractor point which is used to determine the time-frequency assignment of the speaker. We propose three methods for finding the attractors for each source in the embedding space and compare their advantages and limitations. The objective function for the network is standard signal reconstruction error which enables end-to-end operation during both training and test phases. We evaluated our system using the Wall Street Journal dataset (WSJ0) on two and three speaker mixtures and report comparable or better performance than other state-of-the-art deep learning methods for speech separation.
\end{abstract}



\section{Introduction}
\label{sec:intro}

Listening to an individual in crowded situations often takes place in the presence of interfering speakers. Such situations require the ability to separate the voice of a particular speaker from the mixed audio signal of others. Several proposed systems have shown significant performance improvement on the separation task when prior information of speakers in a mixture is given \cite{zhang2016deep, du2016regression}. This however is still challenging when no prior information about the speakers is available, a problem known as speaker-independent speech separation. Humans are particularly adept at this task, even in the absence of any spatial separation between speakers \cite{brungart2001informational, hawley2004benefit, mesgarani2012selective}. This effortless task for humans, however, has proven difficult to model and emulate algorithmically. Nevertheless, it is a challenge that must be solved in order to achieve robust performance in speech processing tasks. For example, while the performance of current automatic speech recognition (ASR) systems has reached that of humans in clean conditions \cite{xiong2016asr}, these systems are still unable to perform well in noisy and crowded environments, lacking robustness when interfering speakers are present. This becomes even more challenging when separating all sources in a mixture is required, such as in meeting transcription and music separation. When signals from multiple microphones are available, beamforming algorithms can be used to improve the target-to-masker ratio \cite{Dahl2012multi, toroghi2012multi}; when only one microphone is available, however, the general problem of audio separation remains largely unresolved. 

Prior to the emergence of deep learning, three main categories of algorithms were proposed to solve the speech separation problem: \textit{statistical} methods, \textit{clustering} methods, and \textit{factorization} methods, with focus on different target tasks. In \textit{statistical} methods, the target speech signal is modeled with probability distributions such as complex Gaussian \cite{simon2009gmm} or methods such as independent component analysis (ICA) \cite{choi2005blind}, where the interference signal is assumed to be statistically independent from the target speech. Maximum likelihood estimation method is typically applied based on the known statistical distributions of the target. In \textit{clustering} methods, the characteristics of the target speaker, such as pitch and signal continuity are estimated from the observation and used to separate the target signal from other sources in the mixture. Methods such as computational auditory scene analysis (CASA) \cite{hu2010tandem, hu2013unsupervised} and spectral clustering \cite{bach2006learning} fall into this category \cite{deliang2006casa}. \textit{Factorization} models, such as non-negative matrix factorization (NMF) \cite{lee2001nmf, virtanen2007monaural, smaragdis2007convolutive}, formulate the separation problem as a matrix factorization problem in which the time-frequency (T-F) representation of the mixture is factorized into a combination of basis signals and activations. The activations learned for each basis signal are then used to reconstruct the target sources. 

In recent years, deep learning has made important progress in audio source separation. Specifically, neural networks have been successfully applied in speech enhancement and separation \cite{xu2013se,zhuo2015se,takkaki2015se, williamson2017time, wang2013towards, huang2015joint, chen2017deep} and music separation \cite{huang2014music, yi2017music} with significantly better performance than that of traditional methods. A typical paradigm for neural networks is to directly estimate T-F masks of the sources given the T-F representation of the audio mixture (such as noisy speech or multiple speakers) \cite{wang2013towards, huang2015joint, hershey2016dpcl, yu2017permutation}. This formulates the separation as a supervised single-class or multi-class regression problem. Different types of masks and objective functions have been proposed. For instance, phase-aware masks for enhancement and separation have been studied in \cite{williamson2017time, hakan2015complex, mayer2017impact}.

Limitations of the previous neural networks become evident when one considers the problem of separating two simultaneous speakers with no prior knowledge of the speakers (speaker-independent scenario). Two main challenges in this situation are the so-called \textit{permutation} problem and the \textit{output dimension mismatch} problem. \textit{Permutation} problem \cite{hershey2016dpcl} refers to the observation that the order of the speakers in the target may not be the same as the order of the speakers in the output of the network. For example, when designing the target output by separating speakers $S_1$ and $S_2$, both $(S_1, S_2)$ and $(S_2, S_1)$ are acceptable permutations of the speakers. Once the target permutation is fixed, however, the output of the network must follow the permutation of the target. In situations where the separation is successful but the outputs have incorrect permutation compared with the targets, the output error will be large, causing the network to diverge from the correct solution.  Aside from this issue, the \textit{output dimension mismatch} problem is also a notable problem. Since the number of speakers in a mixture can vary, a neural network with a fixed number of output targets does not have the flexibility to separate the mixtures where the number of speakers is not equal to the number of output targets. 


Two deep learning methods, deep clustering (DPCL) \cite{hershey2016dpcl} and permutation invariant training (PIT) \cite{yu2017permutation}, have been proposed recently to resolve these problems. In deep clustering, a network is trained to generate a discriminative embedding for each T-F bin so that the embeddings of the T-F bins that belong to the same speaker are closer to each other. Because deep clustering uses the Frobenius norm between the affinity matrix of embeddings and the affinity matrix of ideal speaker assignment (e.g. ideal binary mask) as the training objective, it solves the permutation problem due to the permutation-invariant property of affinity matrices. The mask estimation process is done by applying clustering algorithms such as K-means \cite{hartigan1979algorithm} or spectral clustering \cite{ng2002spectral} to the embeddings, while the assignment of the embeddings to the clusters forms the final estimation. Hence, the number of outputs is only determined by the number of target clusters. While DPCL is able to solve both the permutation and output dimension mismatch problems and produce a state-of-the-art performance, it is unable to use reconstruction error as the target for optimization. This is because the mask generation is done through a post-clustering step on the embeddings, which is done separately from the network network. In more recent DPCL work, minimizing the separation error is processed with an unfolded soft clustering subsystem for direct mask generation, and an additional mask enhancement network is followed for better performance \cite{isik2016single}. 
The PIT algorithm solves the permutation problem by first calculating the training objective loss for all possible permutations for $C$ mixing sources ($C!$ permutations), and then using the permutation with the lowest error to update the network. It solves the output dimension problem by assuming a maximum number of sources in the mixture and using null output targets (very low energy Gaussian noises) as auxiliary targets when the actual number of sources in the mixture is smaller than the number of outputs in the network. PIT was proposed in \cite{yu2017permutation} in a frame-wise fashion and was later shown to have comparable performance to DPCL \cite{kolbaek2017multitalker} with a deep LSTM network structure.

We address the general source separation problem with a novel deep learning framework which we call the `attractor network'. The term ``attractor'' refers to the well-studied effects in human speech perception which suggest that  biological neural networks create perceptual attractors (magnets). These attractors warp the acoustic feature space to draw in the sounds that are close to them, a phenomenon that is called the Perceptual Magnet Effect \cite {kulh1991attr, iverson1995mapping, kuhl2008phonetic}. Our proposed model works on the same principle as DPCL by first generating a high-dimensional embedding for each T-F bin. We then form a reference point (attractor) for each source in the embedding space that pulls all the T-F bins belonging to that source toward itself. This results in the separation of sources in the embedding space. A mask is estimated for each source in the mixture using the similarity between the embeddings and each attractor. Since the correct permutation of the masks is directly related to the permutation of the attractors, our network can potentially be extended to an arbitrary number of sources without the permutation problem once the order of attractors is established. Moreover, with a set of auxiliary points in the embedding space, known as the anchor points, our framework can directly estimate the masks for each source without needing a post-clustering step as in \cite{hershey2016dpcl} or a clustering subnetwork as in \cite{isik2016single}. This aspect creates a system that directly generates the reconstructed spectrograms of the sources in both training and test phases.

The rest of the paper is organized as follows. In Section~\ref{sec:review}, we introduce the general problem of source separation and the embedding learning method. In Section~\ref{sec:model}, we describe the original deep attractor network proposed in \cite{chen2017deep}. In Section~\ref{sec:adanet}, we propose several variants and extensions to the original deep attractor network for better performance. In Section~\ref{sec:exp}, we evaluate the performance of the attractor network and analyze the properties of the embedding space.



\section{Source separation and embedding learning}
\label{sec:review}
We start with defining the general problem of single-channel speech separation, and describe how the method of embedding learning can be used to solve this problem.

\subsection{Single-channel speech separation}

The problem of single-channel speech separation is defined as estimating all the $C$ speaker sources $s_1(t), \ldots, s_c(t)$ given the mixture waveform signal $x(t)$
\begin{align}
x(t) = \sum_{i=1}^C s_i(t)
\label{eqn:prob}
\end{align}

In time-frequency (T-F) domain, the complex short-time Fourier transform (STFT) spectrogram of the mixture, $\mathcal{X}(f,t)$ equals to the sum of the complex STFT spectrograms of all the sources
\begin{align}
\mathcal{X}(f,t) = \sum_{i=1}^C \mathcal{S}_i(f,t)
\label{eqn:stft}
\end{align}

Many speech separation systems use the real-valued magnitude spectrogram  as the input and estimate a set of time-frequency masks for the sources. We denote the flattened magnitude spectrogram, $\lvert\mathcal{X}(f,t)\rvert$, as a feature vector $\vec{x} \in \mathbb{R}^{1\times FT}$, where $F$ is the number of frequency channels and $T$ is the total duration of the utterance. The flattened magnitude spectrograms, $\lvert \mathcal{S}_i(f,t) \rvert$, and corresponding time-frequency masks for source $i = 1, 2, \dotsc, C$ are the vectors $\vec{s}_i \in \mathbb{R}^{1\times FT}$ and $\vec{m}_i \in \mathbb{R}^{1\times FT}$ respectively. The estimated magnitude spectrograms of the source $i$ is denoted by $\hat{\vec{s}}_i \in \mathbb{R}^{1\times FT}$, and calculated by
\begin{align}
\hat{\vec{s}}_i = \vec{x} \odot \vec{m}_i
\end{align}
subject to
\begin{align}
\sum_{i=1}^C \vec{m}_i = \vec{1}
\label{eqn:est}
\end{align}
where $\odot$ is element-wise multiplication and $\vec{1} \in \mathbb{R}^{1\times FT}$ denotes all-one vector. Commonly used masks include ideal binary mask (IBM) \cite{li2009optimality}, ideal ratio mask (IRM) \cite{narayanan2013ideal}, and `wiener-filter' like mask (WFM) \cite{erdogan2015phase}:
\begin{align}
\begin{cases}
IBM_{i,ft} = \delta(\left |\vec{s}_{i,ft}\right | > \left |\vec{s}_{j,ft}\right |),\quad \forall\, j \neq i\\\\
IRM_{i,ft} = \frac{\left |\vec{s}_{i,ft}\right |}{\sum_{j=1}^C \left |\vec{s}_{j,ft}\right |} \\\\
WFM_{i,ft} = \frac{\left |\vec{s}_{i,ft}\right |^2}{\sum_{j=1}^C \left |\vec{s}_{j,ft}\right |^2}
\label{eqn:allmasks}
\end{cases}
\end{align}
where $\delta(x)=1$ if expression $x$ is true and $\delta(x)=0$ otherwise. The reconstruction of the time-domain signals is done by calculating the inverse short-time Fourier transform using the estimated magnitude spectrograms $\hat{\vec{S}}_i$ and the phase of the mixture spectrogram, $\angle {}\mathcal{X}(f,t)$.

\subsection{Source separation in embedding space}

High-dimensional embedding is a powerful and commonly used method in many tasks such as natural language processing and manifold learning \cite{bengio2003neural, mikolov2013distributed, roweis2000nonlinear, maaten2008visualizing}. This technique maps the signal into a high-dimensional space where the resulting representation has desired properties. For example, word embedding is currently one of the standard tools to extract the relationship and connection between different words, and serves as a front-end to more complex tasks such as machine translation and dialogue systems \cite{cho2014learning, serban2016building}. In the problem of source separation in T-F domain, a high-dimensional embedding for each T-F bin is found and speech separation is formulated as a source segmentation problem in the embedding space \cite{hu2013unsupervised, bach2006learning}. 

The embeddings can be either knowledge-based or data-driven. CASA is a popular frameworks for using specially designed features to represent the sources \cite{ellis1996prediction}, where different types of acoustic features are concatenated to produce a high-dimensional embedding which represents the sound source in different time-frequency coordinates. An example of the data driven embedding approach for speech separation is the recently proposed deep clustering method (DPCL) \cite{hershey2016dpcl}. DPCL uses a neural network model to learn embeddings of T-F bins such that to minimize the in-class similarity, while at the same time maximizes the between-class similarity. For the embeddings whose corresponding T-F bins belong to the same speaker, the similarity between them should be large and vice versa. This method therefore creates a high-dimensional representation of the mixture audio that results in better segmentation and separation of speakers. 


\section{Deep Attractor Network}
\label{sec:model}
\begin{figure}[!t]
    \centering
    \includegraphics[width=\columnwidth]{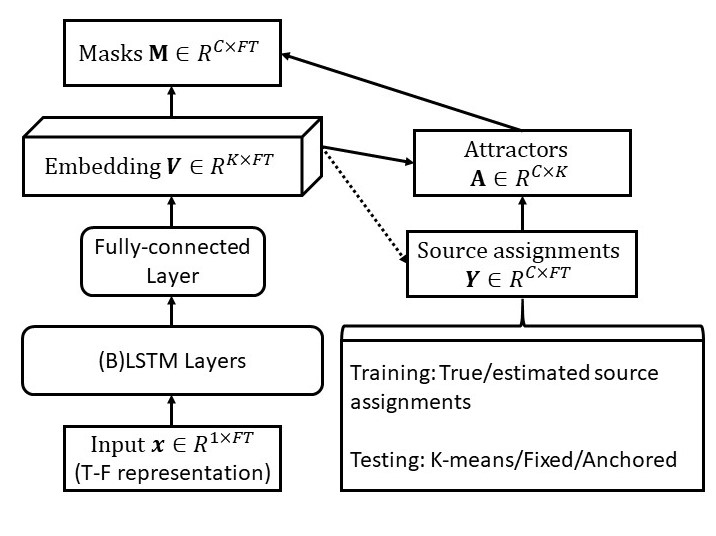}
    \caption{The architecture of DANet. The mixture audio signal is projected into a high-dimensional embedding space. The embeddings of time-frequency bins of each speaker are pulled toward reference points called the attractors. During the training phase, the attractors are formed using true or estimated (in anchored DANet) speaker assignments. During the test phase, the attractors are formed in three alternative ways using: unsupervised clustering, fixed points, and anchor points.}
    \label{fig:sys}
\end{figure}

\begin{figure*}[!t]
    \centering
    \includegraphics[width=17cm]{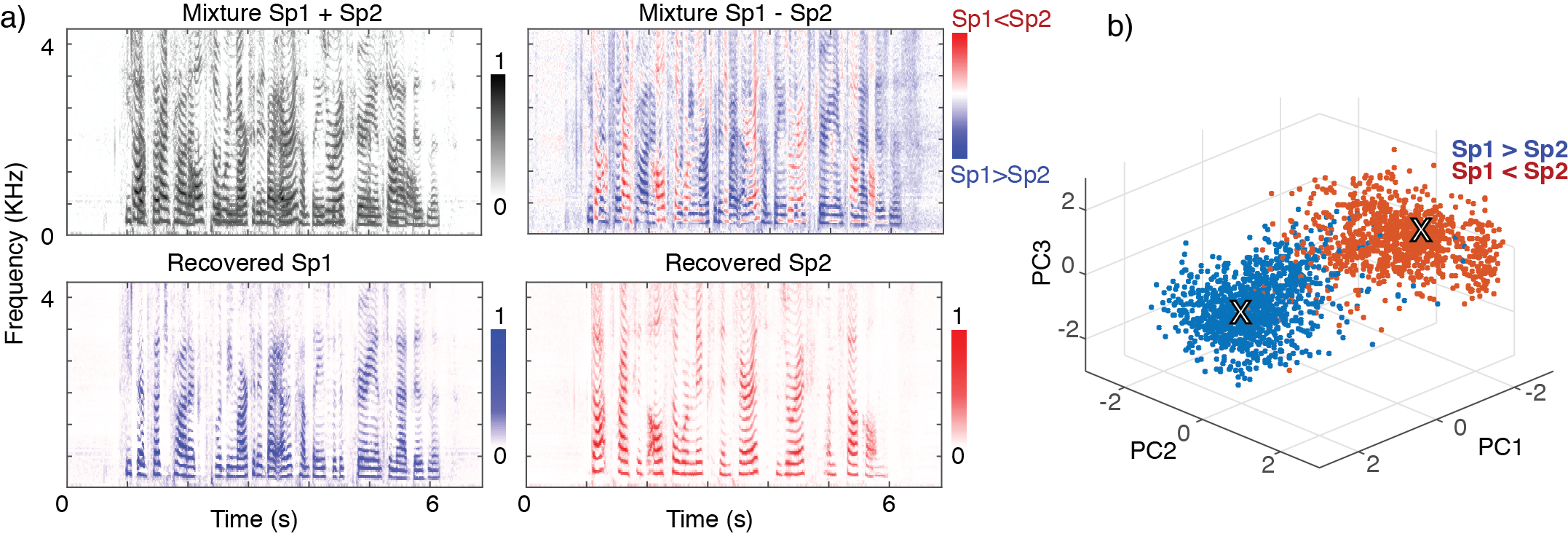}
     \vspace{-0.4cm}
    \caption{a) An example of the mixture spectrogram and the recovered speakers. b) Location of T-F bins in embedding space. Each dot visualizes the first three principle components of the embedding of one T-F bin, where colors distinguish the relative power of speakers in that bin. Attractor points are marked with an X.}
    \label{fig:spec}
\end{figure*}

In this section, we introduce the deep attractor network (DANet) \cite{chen2017deep} and compare it to DPCL \cite{hershey2016dpcl} and PIT \cite{kolbaek2017multitalker}. We then discuss three methods for estimating the parameters of the network during test phase and discuss their pros and cons. We further address the limitation of DANet \cite{chen2017deep} by presenting a new framework called ADANet. Figure~\ref{fig:sys} shows the flowchart of the overall system. Note that in this section we use the same notations as in Section~\ref{sec:review}.

\subsection{Model definition}

Following the main concept of embedding learning with neural network, DANet generates a $K$-dimensional embedding vector for each of the T-F bins in the mixture magnitude spectrogram
\begin{align}
 \vec{V} &= f(\vec{x})
 \label{eqn:emb}
\end{align}
where $\vec{V}\in \mathbb{R}^{K\times FT}$ is the embedding matrix containing all the $K$-dimensional embeddings for the T-F bins and $f$ is the mapping function implemented by the neural network. Instead of using affinity matrices to measure the similarity between embeddings \cite{hershey2016dpcl}, DANet uses reference points (\textit{attractors}) in the embedding space to calculate similarity as well as for mask estimation.

An \textit{attractor} $\vec{a}_i \in \mathbb{R}^{1\times K}$ is a vector in the embedding space that represents a specific speaker, such that all the T-F bins belonging to that speaker are pulled toward the corresponding attractor. The attractors are formed during the training phase by calculating the weighted average of embeddings that belong to each speaker:
\begin{align}
 \vec{a}_{i}&=\frac{\vec{y}_{i}\vec{V}^{\top} }{\sum_{f,t} \vec{y}_{i}},\quad i = 1, 2, \dotsc, C
 \label{eqn:att}
\end{align}
where $\vec{y}_i \in \mathbb{R} ^{1\times FT}$ denotes the speaker assignment for each T-F bin, which can be either the IBM or IRM in this case (Eqn.~\ref{eqn:allmasks}). Since the attractors represent the centroid of each speaker in the embedding space, averaging over the embeddings that correspond to the most salient T-F bins (i.e. the T-F bins with highest power) may lead to a more robust estimation of the attractors. We therefore apply a threshold to the mixture power spectrogram and create a binary threshold vector $\vec{w} \in R^{1\times FT}$ that filters out the T-F bins with low power. Given a parameter $\rho$, the threshold vector $\vec{w}$ is defined as 
\begin{align}
\vec{w}_{ft} &= 
\begin{cases}
1, \quad \text{if}\quad \vec{x}_{ft} > \rho\\
0, \quad \text{else}
\end{cases}
 \label{eqn:threshold}
\end{align}
The attractors are then estimated as follows:
\begin{align}
\vec{a}_i&=\frac{(\vec{y}_{i} \odot \vec{w})\vec{V}^\top }{\sum_{f,t} (\vec{y}_{i} \odot \vec{w})},\quad i = 1, 2, \dotsc, C
 \label{eqn:att_weight}
\end{align}

After the generation of the attractors, DANet calculates the similarity between the embedding of each T-F bin and each attractor:
\begin{align}
\vec{d}_{i} &= \vec{a}_{i} \vec{V},\quad i = 1, 2, \dotsc, C
\label{eqn:distance}
\end{align}
where $\vec{d}_i\in \mathbb{R}^{1\times FT}$ denotes the distance of each T-F bin in the embedding space from the attractor $i$. This distance $\vec{d}_i$ is small for the embeddings that are close to an attractor, and is large for points that are far away. The mask for each speaker $\hat{\vec{m}}_i \in \mathbb{R}^{1\times FT}, i = 1, \dotsc, C$ is then estimated by normalizing the similarity distance using a nonlinear function to constrain the mask to the range $[0, 1]$
\begin{align}
\hat{\vec{m}}_{i}&=\mathcal{H}(\vec{d}_{i}),\quad i = 1, 2, \dotsc, C
\label{eqn:mask}
\end{align}
where $\mathcal{H}$ is a nonlinear function which can be either the Softmax or Sigmoid function that is applied to each element of $\vec{d}_i$:
\begin{align}
\begin{cases}
Softmax(\vec{d}_{i,ft}) = \frac{e^{\vec{d}_{i,ft}}}{\sum_{i=1}^C e^{\vec{d}_{i,ft}}}\\\\
Sigmoid(\vec{d}_{i,ft}) = \frac{1}{\sum_{i=1}^C (1+e^{-\vec{d}_{i,ft}})}
\end{cases}
\label{eqn:activation}
\end{align}

The neural network is then trained by minimizing a standard $L^2$ reconstruction error as the objective function
\begin{align}
{l}&=\frac{1}{C}\sum_{i}\left \| \vec{x} \odot (\vec{m}_{i} - \hat{\vec{m}}_{i})\right \|^2_2
\label{eqn:err}
\end{align}
where $\vec{m}_i$ and $\hat{\vec{m}}_i$ $\in \mathbb{R}^{1\times FT}$ are the ideal and estimated target masks for the speakers. Since the masks are a function of both attractors and embeddings, optimizing the reconstruction error (Eqn.~\ref{eqn:err}) forces the network to pull the embeddings that belong to the same speaker together and place the attractors far from each other. The initial embeddings and attractors are randomly distributed in the embedding space, and as the training continues, the attractors gradually become separated and pull the embeddings to a proper space that can separate different speakers. 

The permutation of the speakers in the outputs is determined by the permutation of the attractors in the embedding space. This is further determined by the permutation of the given speaker assignment vectors $\vec{y}_i$. Hence, once the permutation of the target masks $\vec{m}_i$ matches the permutation of speaker assignment function $\vec{y}_i$, we no longer have the permutation problem in DANet. On the other hand, since the number of the speakers is determined by the number of attractors which is a function of $\vec{y}_i$, the output dimension problem is solved in DANet framework, since the number of attractors can change dynamically without changing the network structure.

\subsection{Relation to DPCL and PIT}

Since DANet shares the same network structure as DPCL \cite{hershey2016dpcl}, it is important to illustrate the difference between DANet and DPCL. In contrast to DPCL, DANet directly optimizes the reconstruction error with a computationally simpler objective function rather than the calculation of affinity matrices in DPCL. Moreover, the direct mask estimation allows it to use flexible similarity measurements and target masks, including phase-aware  \cite{gaich2015phase} and phase-sensitive mask \cite{erdogan2015phase}.

On the other hand, when the attractors are considered as the trained weights of the network instead of dynamically formed by the embeddings (Eqn. ~\ref{eqn:att} \&~\ref{eqn:att_weight}), DANet reduces to a classification network \cite{zhuo2015se,xu2013se} and equation~\ref{eqn:distance} is equivalent to a linear fully-connected layer. In this case, permutation invariant training becomes necessary since the masks are no longer linked to a speaker. However, the dynamic formation of the attractors in DANet allows utterance-level flexibility in the generation of the embeddings. Moreover, DANet does not assume a fixed number of outputs, since the number of attractors is decided by the size of the speaker assignment function $Y$ during training. 

\section{Estimation of the attractor points}
\label{sec:adanet}
As described in equations~\ref{eqn:att} \&~\ref{eqn:att_weight}, the actual speaker assignment (e.g. using the IBM or IRM methods, Eqn.~\ref{eqn:allmasks}) is necessary to form the attractors during the training phase. However, this information is not available during the test phase, causing a mismatch between training and test phases. In this section, we propose several methods for estimating the location of the attractors during the test phase. The first two methods were introduced in \cite{chen2017deep}. Here we discuss their limitation and propose a new method for estimating the attractor points called Anchored DANet (ADANet), an extension that enables direct attractor estimation and mask generation for both training and test phases.

\subsection{Forming the attractors using clustering}
\label{sec:clustering}

The simplest method to form the attractors in the embedding space is to use an unsupervised clustering algorithm such as K-means on the embeddings to determine the speaker assignment (DANet-Kmeans in Fig.~\ref{fig:anchor-flow} and Section~\ref{sec:exp}). This method is similar to \cite{hershey2016dpcl}. In this case, the centers of the clusters are treated as the attractors for mask generation. Figure~\ref{fig:spec} shows an example of this method, where the crosses represent the centers of the two clusters, which are also the estimated attractors.

\subsection{Fixed attractor points}
\label{sec:fixed}

\begin{figure}[!t]
    \centering
    \includegraphics[width=8cm]{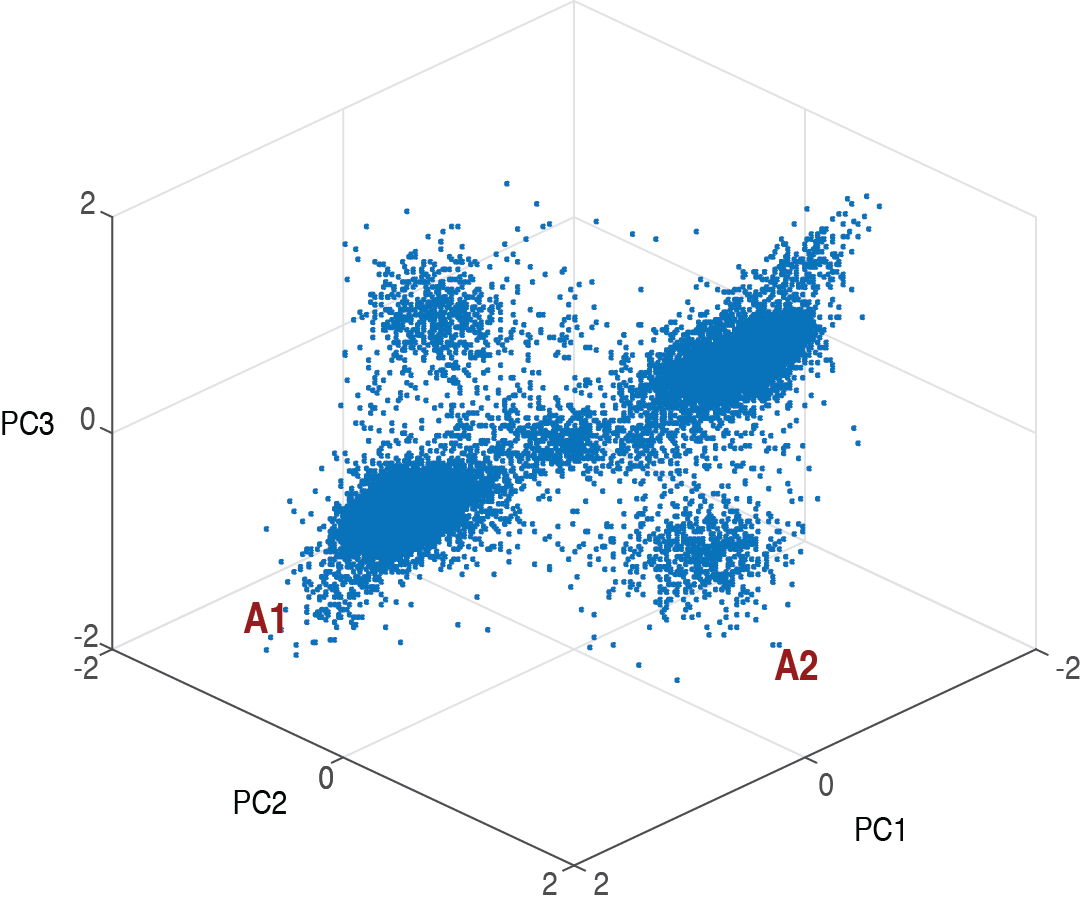}
    \caption{Location of attractor points in the embedding space in a sigmoid network. Each dot corresponds to one of the 10000 mixtures sounds, visualized using the first three principal components. Two distinct attractor pairs are visible (denoted by A1 and A2).}
    \label{fig:attractor_all}
\end{figure}

While there is no direct constraint on the location of the attractors in the embedding space, we have found empirically that the location of the attractor points are relatively constant across different mixtures. Figure~\ref{fig:attractor_all} shows the location of attractors for 10000 different mixtures, with each opposite pair of dots corresponding to the attractors for the two speakers in a given mixture. Two pairs of attractors (marked as A1 and A2) are automatically discovered by the network. Based on this observation, we propose to first estimate all the attractors in the training phase for different mixtures, and subsequently use the mean of those attractors as the fixed attractors during the test phase (DANet-Fixed in Fig.~\ref{fig:anchor-flow} and Section~\ref{sec:exp}). The advantage of using fixed attractors is that it removes the need for the clustering step, allowing the system to directly estimate the mask for each time frame and enabling real-time implementation. 

\subsection{Anchored DANet (ADANet)}
\begin{figure}[!tbp]
    \centering
    \includegraphics[width=8cm]{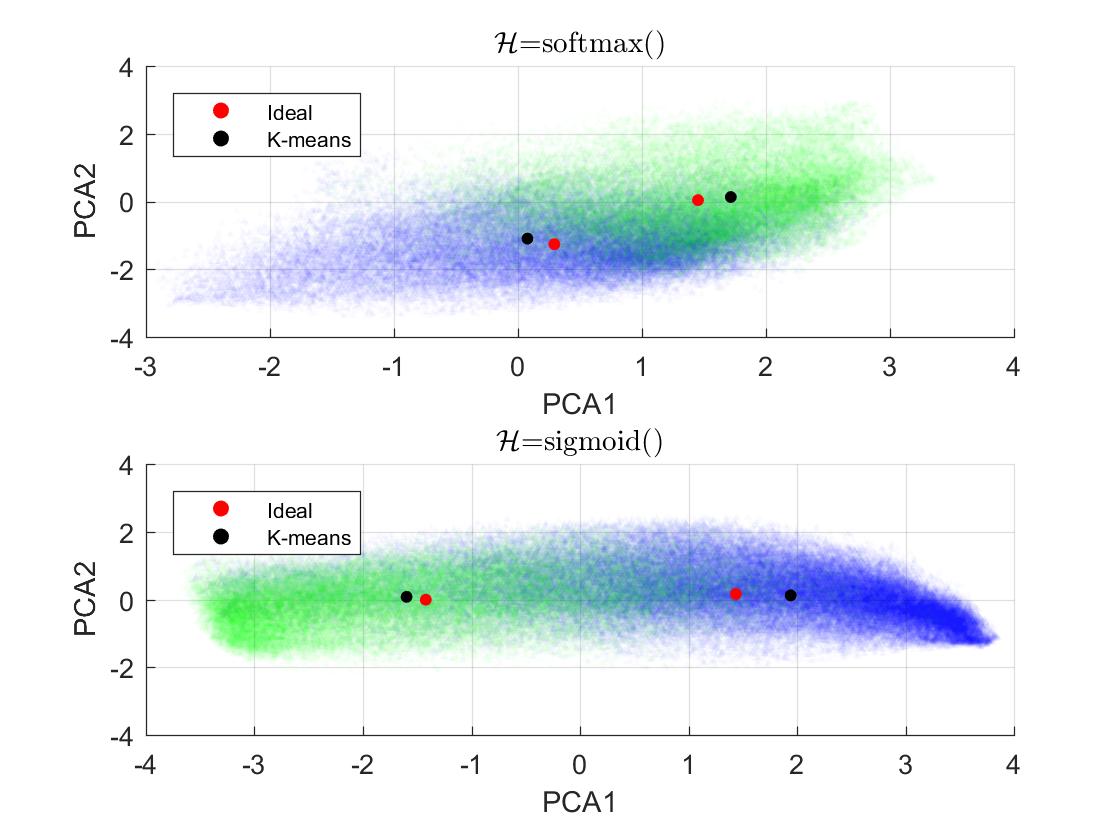}
    \caption{The gap between the true location of attractors (red dots) and the estimated location using K-means clustering (black dots) in the embedding space, for two networks with softmax and sigmoid nonlinearities. Blue and green dots show the PCA projection of the embedding of T-F bins corresponding to speakers one and two.}
    \label{fig:mismatch}
\end{figure}

While both clustering and fixed attractor method can be used during the test phase, these approaches have several limitations. For clustering-based estimation, the K-means step increases the computational cost of the system and therefore increases the run-time delay. Additionally, the centers of the clusters are not guaranteed to match the true locations of the attractors as the (weighted) averages of the corresponding embeddings. This potential difference between the true and estimated attractors causes a mismatch in the mask formation between the training and test phases. One such instance is shown in figure~\ref{fig:mismatch}, where the embedding space is visualized using its first two principle components. The locations of true and estimated attractors are plotted in yellow and black, and the distance between the two shows the mismatch between the true and estimated attractor locations. As suggested by equation~\ref{eqn:mask}, this mismatch changes the mask estimated for each source and thus reduces the accuracy of the separation. We refer to this problem as the \textit{center mismatch problem}, which is caused by the unknown speaker assignment during the test phase. Using fixed attractors on the other hand relies on the assumption that the location of the attractors in training and test phases are similar. This, however, may not be the case if the test condition is significantly different from the training condition. Another drawback of fixing the number and location of attractors is the lack of flexibility when dealing with mixtures with variable number of speakers. 

\begin{figure*}[!t]
    \centering
    \includegraphics[width=17cm]{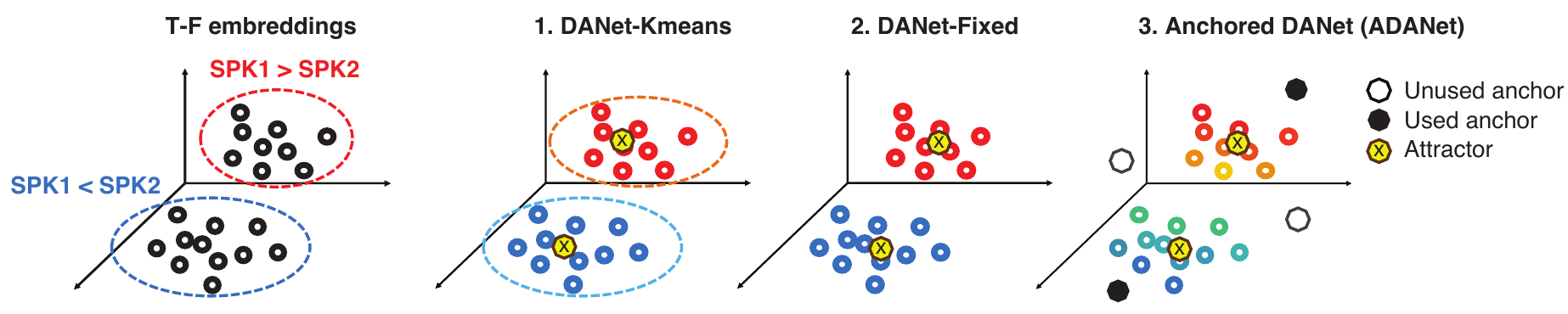}
    \caption{Different methods for estimating the attractors in DANet. During the test phase, the attractors can be found in three ways: 1) by  clustering the embeddings, 2) using fixed attractors calculated from the training set, and 3) first using anchor points to calculate the speaker assignments, followed by the attractor estimation given the speaker assignments.}
    \label{fig:anchor-flow}
\end{figure*}

To remedy this problem, we propose the Anchored DANet (ADANet), in which several trainable reference points in the embedding space (anchors) are used to estimate the speaker assignment $\vec{Y}$ in both training and test phases (Fig.~\ref{fig:anchor-flow}). The estimated assignments are then used to find the attractors. This removes the need for true speaker assignment also during the training, and thus removes the \textit{center mismatch problem}.

\begin{figure*}[!ht]
    \centering
    \includegraphics[width=17cm]{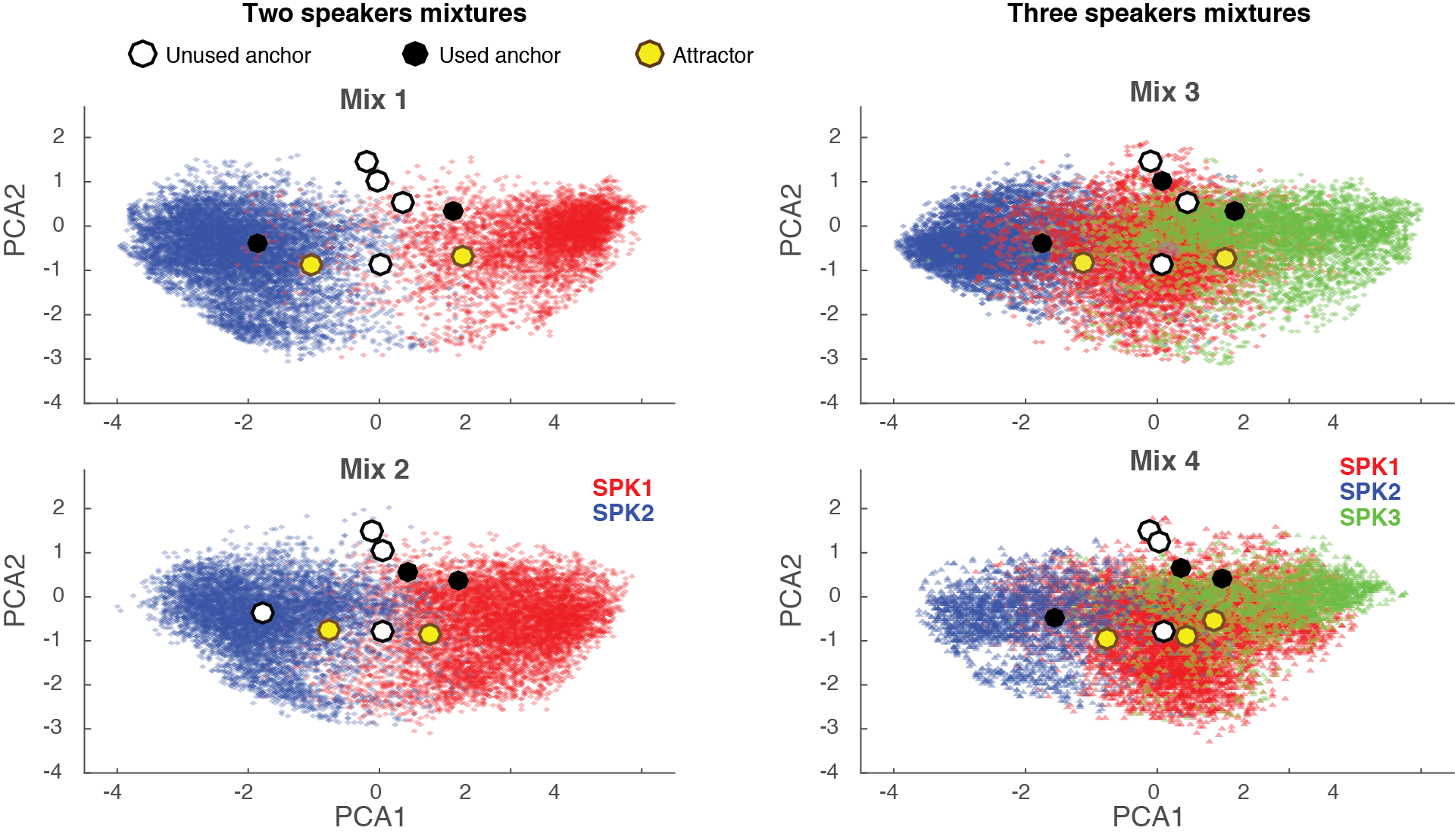}
    \caption{The PCA projection of the embeddings of the speakers (SPK1..3), anchor points, and attractor points (yellow) for two different two-speaker mixtures, and two different three-speaker mixtures in a single six-anchor point DANet. As can be seen, different anchor points are selected for different mixtures, illustrating the flexibility and stability of ADANet when dealing with different mixture conditions.}
    \label{fig:indicators}
\end{figure*}

Similar to the original DANet \cite{chen2017deep}, a $K$-dimensional embeddings $\vec{V}$ for the T-F bins are first generated (Eqn.~\ref{eqn:emb}). ADANet first creates $N$ randomly initialized, trainable points in the embedding space $\vec{V}$, which are denoted by $\vec{b}_j \in \mathbb{R}^{1\times K},\, j = 1, 2, \dotsc, N$ , which we call \textit{anchor points}. The number of anchors $N$ is chosen to be no smaller than the number of speakers in the mixture signals in the training set, $C$. This ensures that the network will have enough capacity to deal with different number of speakers in one general network. In a mixture which contains $C$ speakers, we first choose all the $C$ combinations of the $N$ anchor points $[\vec{b}_1, \vec{b}_2, \dotsc,  \vec{b}_N]$, resulting in total of $\binom{N}{C}$ combinations denoted by $\vec{L}_p \in \mathbb{R}^{C\times K},\, p = 1, 2, \dotsc, \binom{N}{C}$, where $\binom{N}{C}$ is the standard binomial coefficient. We then find the distance of the embeddings from the anchor points in each subset $\vec{L}_p$, and use it to estimate the speaker assignment: 
\begin{gather}
  {\vec{D}_p} = \vec{L}_{p}\vec{V}, \quad p = 1, 2, \dotsc, \binom{N}{C} \\
  \hat{\vec{Y}}_{p} = Softmax({\vec{D}_p})
  \label{eqn:fix_weight}
\end{gather}
where ${\vec{D}}_p \in \mathbb{R}^{C\times FT}$ is the distance between embeddings and each of the $C$ anchors in subset $p$, and ${\hat{\vec{Y}}_p} \in \mathbb{R}^{C\times FT}$ is the estimated speaker assignment for the corresponding subset. The Softmax function (Eqn.~\ref{eqn:activation}) is used to increase the dynamic range of the weights so that they are pushed closer to 0 or 1. The $C$ attractors for each anchor subset are then calculated using the estimated speaker assignment according to equation~\ref{eqn:att} or~\ref{eqn:att_weight}, leading to $\binom{N}{C}$ sets of attractors $\vec{A}_p\in \mathbb{R}^{C\times K}$. The set with the minimum in-set similarity (i.e. largest in-set distance between attractors) is selected for mask estimation
\begin{gather}
 \vec{S}_p = \vec{A}_{p}\vec{A}_{p}^\top, \quad \vec{S_p}  \in \mathbb{R}^{C\times C}\\[2ex]
 s_p = max\{(\vec{s}_{p_{ij}})\},\quad i\neq j\\[2ex]
 \hat{\vec{A}} = \argmin_{\vec{A}_p}\{s_p\}, \quad p = 1, 2, \dotsc, \binom{N}{C}
\label{eqn:select}
\end{gather}
where $s_p$ is a scalar that represents the maximum similarity between any two attractors in $\vec{A}_p$, and ${\vec{A}}\in \mathbb{R}^{C\times K}$ is the set of attractors with smallest in-set similarity among all $\binom{N}{C}$ attractor sets. Given the attractors $\vec{A}$, the mask estimation is done in the same manner as equation~\ref{eqn:distance} \&~\ref{eqn:mask}. Figure~\ref{fig:indicators} shows examples of the position of the embeddings and the anchor points in a six-anchor point network for four different mixture conditions (two two-speaker mixtures and two three-speaker mixtures), where different sets of anchor points are chosen for estimation of attractors in different conditions. Note that the same network is used for all the mixtures, and training of the network is described in Section~\ref{sec:exp}.

The entire process of ADANet can be represented as a generalized Expectation-Maximization (EM) framework, where the speaker assignment calculation is the ``Expectation step'' and the following attractor formation step is the ``Maximization step.'' Hence, the separation procedure of ADANet can be viewed as a single EM iteration. Note that this method is similar to the unfolded soft-clustering subnetwork proposed in \cite{isik2016single} but the parameters (i.e. statistics) for the clusters here are dynamically defined by the embeddings. Thus, they do not increase the total number of parameters in the network. Moreover, this approach also allows utterance-level flexibility in the estimation of the clusters.

\vspace{-0.1cm}

\subsection{Attractor formation summary}

Compared to the clustering and fixed attractor methods, ADANet eliminates the need for true speaker assignment during both training and test phases. The \textit{center mismatch problem} no longer exists because the attractors are determined by the anchor points, which are trained in the training phase and fixed in the test phase. The mask generation process is therefore matched during both training and test phases. Figure~\ref{fig:anchor-flow} shows the difference between various ways of attractor calculation. The fixed attractor method enables real-time processing and generates the masks without the extra clustering step, but is sensitive to the mismatch between training and test conditions. ADANet framework solves the \textit{center mismatch problem} and allows direct mask generation in both training and test phase, however it increases the computational complexity since all $\binom{N}{C}$ subsets of anchors require one EM iteration. Moreover, since the correct permutation of the anchors is unknown, permutation invariant training in \cite{yu2017permutation} is required for training the ADANet.


\section{Experiments and analysis}
\label{sec:exp}
We evaluate our proposed model on the task of single-channel two and three speaker separation. Example sounds can be found here \cite{web2016danet}.

\subsection{Data}

We use the WSJ0-2mix and WSJ0-3mix datasets introduced in \cite{hershey2016dpcl} which contains two 30 h training sets and a 10 h validation sets for the two tasks. These tasks are generated by randomly selecting utterances from different speakers in the Wall Street Journal (WSJ0) training set si\_tr\_s and mixing them at various signal-to-noise ratios (SNR) randomly chosen between 0 dB and 5 dB. Two 5 h evaluation sets are generated in the same way, using utterances from 16 unseen speakers from si\_dt\_05 and si\_et\_05 in the WSJ0 dataset. All data are resampled to 8 kHz to reduce computational and memory costs. The log magnitude spectrogram serves as the input feature, computed using short-time Fourier transform (STFT) with 32 ms window length (256 samples), 8 ms hop size (64 samples), and the square root of Hanning window.

\subsection{Evaluation metrics} 
We evaluated the separation performance on the test sets using three metrics: signal-to-distortion ratio (SDR) \cite{vincent2006performance} for comparing with PIT models in \cite{yu2017permutation, kolbaek2017multitalker}, scale-invariant signal-to-noise ratio (SI-SNR) to compare with DPCL models in \cite{isik2016single}, and PESQ score \cite{rix2001perceptual} for evaluation of the speech quality. The SI-SNR, proposed in \cite{isik2016single}, is defined as:
\begin{gather}
    \vec{s}_{target} := \frac{\langle \hat{\vec{s}}, \vec{s} \rangle \vec{s}}{\left \| \vec{s} \right \|^2}\\
    \vec{e}_{noise} := \hat{\vec{s}} - \vec{s}_{target}\\
    \text{SI-SNR} := 10\,log_{10}\frac{\left \| \vec{s}_{target} \right \|^2}{\left \| \vec{e}_{noise} \right \|^2}
    \label{eqn:sdr}
\end{gather}
where $\hat{\vec{s}} \in \mathbb{R}^{1\times t}$ and $\vec{s} \in \mathbb{R}^{1\times t}$ are the estimated and original clean source respectively, $t$ denotes the length of the signals, and $\left \| \vec{s} \right \|^2 = \langle \vec{s}, \vec{s} \rangle$ denotes the power of the signal. $\hat{\vec{s}}$ and $\vec{s}$ are both normalized to have zero-mean to ensure scale-invariance.

\subsection{Network architecture}

The network contains 4 Bi-directional LSTM \cite{hoch1997lstm} layers with 600 hidden units in each layer. The embedding dimension is set to 20 according to \cite{hershey2016dpcl}, resulting in a fully-connected feed-forward layer of 2580 hidden units (20 $\times$ 129 as K $\times$ F) after the BLSTM layers. Adam algorithm \cite{diederik2014adam} is used for training, with the learning rate starting at $1e^{-3}$ and then halved if no best validation model is found in 3 epochs. The total number of epochs is set to 100, and we used the cost function in equation~\ref{eqn:err} on the validation set for early stopping. The criterion for early stopping is no decrease in the loss function on validation set for 10 epochs. WFM (Eqn. \ref{eqn:allmasks}) is used as the training target.

We split the input features into non-overlapping chunks of 100-frame and 400-frame length as the input to the network with a curriculum training strategy \cite{bengio2009curriculum}. We first train the network with 100-frame length input until converged, and then continue training the network with 400-frame long segments. The initial learning rate for 400-frame length training is set to be $1e^{-4}$ with the same learning rate adjustment and early stopping strategies. For comparison, we trained the networks with and without curriculum training. We also studied the effect of dropout \cite{srivastava2014dropout} during the training of the network by which was added with probability of 0.5 to the input to each of the BLSTM layers. During test phase, the number of speakers is provided to both DANet and ADANet in all the experiments except for the mixed number of speaker experiment shown in Table~\ref{tab:mix_spk}.

\subsection{Results}


We first examined the effect of choosing different threshold values for attractor estimation. Table~\ref{tab:2spk_weight} shows the results for different thresholds values (Eqn.~\ref{eqn:threshold}) in attractor formation, as well as the effect of the nonlinearity functions used for mask generation (Eqn.~\ref{eqn:mask}). The 90\% suffix denotes a threshold $\rho$ that keeps the top $90\%$ of T-F bins in the mixture spectrogram in equation~\ref{eqn:threshold}. Notation  `-Kmeans' refers to clustering based attractor formation (Section~\ref{sec:clustering}), and  `-Fixed' indicates networks with fixed attractors (Section~\ref{sec:fixed}). The suffix `-do' indicates training with dropout and the superscript `$^*$' denotes curriculum training.

The results in table~\ref{tab:2spk_weight} show that using the \%90 threshold leads to better performance compared to no threshold, which indicates the importance of accurate estimation of attractors for mask generation. Moreover, we observe that although Softmax with ideal speaker assignment leads to a higher performance than Sigmoid, the performance with K-means for Softmax networks are worse than Sigmoid networks. 
Our results also show that the K-means performance in the Softmax network is highly dependent on how the network is optimized (i.e. different network trained with different initial values may have very different performance) with a similar level of validation error. This observation is supported by our assumption that Softmax is less sensitive to the distance when the embedding dimension $N$ is high. As a result, adding a training regularization such as dropout does not guarantee improved performance for K-means clustering, which is also a disadvantage of a network architecture that does not directly optimize the mask generation.


\begin{table}[!htbp]
\small
\centering
\caption{\small SI-SNR improvement (SI-SNRi) in decibel on WSJ0-2mix with different attractor formation strategies.}
\label{tab:2spk_weight}
\begin{tabular}{c|c|cc}
\thline
Method & $\mathcal{H}$ & K-means & ideal \\
 \thline
DANet & Sigmoid & 9.3 & 9.6 \\
DANet-90\% & Sigmoid & 9.5 & 9.7 \\
DANet-90\%-Fixed & Sigmoid & 9.2 & - \\
DANet-90\%$^*$ & Sigmoid & \bf{10.3} & 10.6\\
DANet-90\% & Softmax & 9.4 & 9.9\\
DANet-90\%$^*$ & Softmax & 10.0 & 10.8 \\
\thline
\end{tabular}
\end{table}


Table~\ref{tab:2spk_indicator} shows the effect of changing the number of anchors in ADANet with Softmax for mask generation. 
All the networks hereafter apply a 90\% threshold for estimation of attractors. As the number of anchors increases, the performance of the network consistently improves. It confirms that the increased flexibility in the choice of the anchors help the final separation performance. Unlike Softmax DANet with K-means, adding dropout in BLSTM layers here always increases the performance. This shows the advantage of the framework without using the post-clustering step.

\begin{table}[!htbp]
\small
\centering
\caption{\small SI-SNR improvements on the WSJ0-2mix in ADANet with varying number of anchor points.}
\label{tab:2spk_indicator}
\begin{tabular}{c|c}
\thline
\small \# of anchors & \small SI-SNRi\\
 \thline
2 &  9.3 \\
4 &  9.5 \\
6 & 9.6 \\
6$^*$ & 10.1 \\
6-do$^*$ & \bf{10.4} \\
\thline
\end{tabular}
\end{table}


Tables~\ref{tab:2spk_compare} \&~\ref{tab:3spk_res} compare our method with other speaker-independent techniques in separation of two and three speaker mixtures. DPCL \cite{hershey2016dpcl} is the original deep clustering model with K-means clustering for mask generation, and DPCL++  \cite{isik2016single} is an extension of deep clustering that directly generates the masks with soft-clustering layers, and further improve the masks with a second stage enhancement network. uPIT-BLSTM \cite{kolbaek2017multitalker} is the utterance level PIT model that applies PIT on the entire utterance using deep BLSTM network, and uPIT-BLSTM-ST contains a tandem second-stage mask enhancement network for performance improvement. The DPCL and DPCL++ methods also applied a similar curriculum training strategy, while the uPIT-BLSTM and uPIT-BLSTM-ST methods only used standard training. In both tables, we categorize the methods into single-stage and two-stage systems for better comparison, where the two-stage systems have a second-stage enhancement network. The ADANet system in this comparison utilizes six anchors.

In WSJ0-2mix test set (Table~\ref{tab:2spk_compare}), the K-means DANet outperforms all the previous one-stage systems and the two-stage uPIT-BLSTM-ST system. The ADANet with six anchors has the best performance among all one-stage systems, and is only slightly worse than the two-stage DPCL++ system. In WSJ0-3mix test set (Table~\ref{tab:3spk_res}), both K-means DANet and six-anchor ADANet outperform all previous systems with either one-stage or two-stage configuration, and ADANet also performs significantly better than K-means DANet.
The PESQ scores for DANet and ADANet in WSJ0-2mix show significant improvement upon mixture, while in WSJ0-3mix the gaps between WFM and the models are larger. This is expected since the speaker with lowest energy is harder to separate than the two speaker cases, which may lead to a lower PESQ score.

\begin{table}[!htbp]
\small
\centering
\caption{\small Comparison of DANet and ADANet with other methods on WSJ0-2mix.}
\label{tab:2spk_compare}
\begin{tabular}{c|c|c|c|c}
\thline
Method & Stages & SI-SNRi & SDRi & PESQ \\
 \thline
DPCL \cite{hershey2016dpcl} & 1 & - & 5.8 & -\\
uPIT-BLSTM \cite{kolbaek2017multitalker} & 1 & -  & 9.4 & - \\
DPCL++ \cite{isik2016single} & 2 &\bf{10.8}  & - & -\\
uPIT-BLSTM-ST \cite{kolbaek2017multitalker} & 2 & -  & 10.0 & - \\
\hline
DANet-Kmeans$^*$ & 1 & 10.0  & 10.3 & 2.64 \\
DANet-Fixed$^*$ & 1 & 9.9  & 10.2 & 2.57 \\
ADANet-6-do$^*$ & 1 & 10.4  & \bf{10.8} & \bf{2.82} \\
\hline
Mixture & - & - & - & 2.01 \\
WFM & - & 13.9 & 14.2 & 3.66 \\
\thline
\end{tabular}

\end{table}

\begin{table}[!htbp]
\small
\centering
\caption{\small Comparison of DANet and ADANet with other methods on WSJ0-3mix.}
\label{tab:3spk_res}
\begin{tabular}{c|c|c|c|c}
\thline
Method & Stages & SDRi & SI-SNRi & PESQ\\
 \thline
uPIT-BLSTM \cite{kolbaek2017multitalker} & 1 & 7.4 & - & -\\
uPIT-BLSTM-ST \cite{kolbaek2017multitalker} & 2 & 7.7 & - & -\\
DPCL++ \cite{isik2016single} & 2 & - & 7.1 & -\\
\hline
DANet-Kmeans$^*$ & 1 & 8.9 & 8.6 & 1.92 \\
ADANet-6-do$^*$ & 1 & \bf{9.4} & \bf{9.1} & \bf{2.16} \\
\hline
Mixture & - & - & - & 1.65 \\ 
WFM & - & 15.0 & 15.3 & 3.40 \\
\thline
\end{tabular}
\end{table}

Table~\ref{tab:mix_spk} shows the results in the speaker separation task where the same network is used to separate both two and three speaker mixtures. The networks are first trained on three-speaker mixtures until convergence, and then continued training using both two- and three-speaker mixtures. For K-means DANet, the information about the number of speakers is given during both training and test phases, and for ADANet, we use a training strategy similar to \cite{kolbaek2017multitalker}, where the network always uses three anchors to generate three masks, and an auxiliary output mask of all zero entries is concatenated to the target masks in the two-speaker cases. During test phase, no information about the number of speakers is provided, and a simple energy-based detector is used to detect the number of speakers. If the power in an output is 20 dB less than the other outputs, it is subsequently discarded.

Table~\ref{tab:mix_spk} shows that K-means DANet performs well on the two-speaker separation task, but has worse performance on three-speaker separation. This may be due to the less separated embeddings in two-speaker cases for Softmax function discussed in Table~\ref{tab:2spk_weight}. However, with the network configuration in ADANet, the performance is significantly better than both one-stage and two-stage PIT systems. Moreover, ADANet successfully detects the correct number of outputs in all the 3000 utterances in the WSJ0-2mix test set, showing that appending a zero-mask to the target masks of two speaker mixtures during training enables the network to learn a silent mask under two-speaker cases without sacrificing the performance in three-speaker separation. This also indicates that ADANet can automatically select proper anchors for speaker assignment estimation in mixtures with different number of sources (Fig.~\ref{fig:indicators}).



\begin{table}[!htbp]
\small
\centering
\caption{\small SDR improvements (dB) for 2 and 3 speaker mixtures trained on both WSJ0-2mix and WSJ0-3mix.}
\label{tab:mix_spk}
\begin{tabular}{c|c|cc}
\thline
Method & Stages & 2 Spk & 3 Spk \\
 \thline
uPIT-BLSTM \cite{kolbaek2017multitalker} & 1 & 9.3 & 7.1 \\
uPIT-BLSTM-ST \cite{kolbaek2017multitalker} & 2 & 10.1 & 7.8 \\
\hline
DANet-Kmeans$^*$ & 1 & 10.2 & 5.3  \\
ADANet-6-do$^*$ & 1 & \bf{10.4} & \bf{8.5} \\
\thline
\end{tabular}
\end{table}


\section{Conclusion}
\label{sec:conclusion}
In this paper, we introduce the deep attractor network (DANet) for single-microphone speech separation. We discussed its advantages and drawbacks, and proposed an extension, Anchored DANet (ADANet) for better and more stable performance. DANet extends the deep clustering framework by creating attractor points in the embedding space which pull together the embeddings that belong to a specific speaker. Each attractor in this space is used to form a time-frequency mask for each speaker in the mixture. Directly minimizing the reconstruction error allows better optimization of the embeddings. We explored the mismatch problem of DANet during training and test phase, which is resolved in ADANet to allow direct mask generation in both training and test phases. This removes the post-clustering step for mask estimation. The ADANet approach provides a flexible solution and a generalized Expectation-Maximization strategy to deterministically locate the attractors from the estimated speaker assignment. This approach maintains utterance-level flexibility in attractor formation, and can also generalize to changing signal conditions. Experimental results showed that compared with previous state-of-art systems, both DANet and ADANet have comparable or better performance in both two- and three-speaker separation tasks.

The future works include exploring ways to incorporate speaker information in the estimation of anchor points and in the generation of the embeddings. These aspects can result in both speaker-dependent and speaker-independent separation by the same network. Separating more sources beyond speech signals by creating a larger and more informative embedding space is also an interesting and important topic, which prompts the possibility of designing a universal acoustic source separation framework.

\section{Acknowledgement}
The authors would like to thank Drs. John Hershey and Jonathan Le Roux of Mitsubishi Electric Research Lab, and Dr. Dong Yu of Tencent AI Lab for constructive discussions. Yi Luo and Zhuo Chen contributed equally to this work. This work was funded by a grant from National Institute of Health, NIDCD, DC014279, National Science Foundation CAREER Award, and the Pew Charitable Trusts.

\vfill\pagebreak

\bibliographystyle{IEEEbib}
\bibliography{Template}

\end{document}